\documentclass[%
 reprint,
 amsmath,amssymb,
 aps,
prb,
]{revtex4-2}

\usepackage{graphicx}% Include figure files
\usepackage{dcolumn}% Align table columns on decimal point
\usepackage{bm}% bold math
\usepackage{todonotes}
\usepackage{hyperref}
\usepackage{soul}
\usepackage{hyperref}
\hypersetup{
    colorlinks=true,
    linkcolor=blue,
    filecolor=magenta,      
    urlcolor=black,
    citecolor=red
    }% add hypertext capabilities
\DeclareUnicodeCharacter{2212}{-}
\usepackage{multirow}
\usepackage{dcolumn}
\usepackage[table]{xcolor}
\usepackage{colortbl}
\newcolumntype{d}[1]{D{.}{.}{#1}}

\begin{document}

\preprint{APS/123-QED}

\title{The role of stacking and strain in mean-field magnetic moments of multilayer graphene}% Force line breaks with \\

\author{András Balogh}
\email{kisbalogh99@student.elte.hu}
\affiliation{ELTE Eötvös Loránd University Department of Biological Physics}

\author{János Koltai}
\affiliation{ELTE Eötvös Loránd University Department of Biological Physics}

\author{Péter Nemes-Incze}
\affiliation{Hungarian Research Network, Centre for Energy Research, Institute of Technical Physics and Materials Science, 1121 Budapest, Hungary}
\affiliation{MTA - HUN-REN EK Lendület "Momentum" Topology in Nanomaterials Research Group, 1121 Budapest, Hungary}

\author{\textcolor{black}{Zoltán Tajkov}}
\affiliation{\textcolor{black}{ELTE Eötvös Loránd University Department of Physics of Complex Systems}}
\affiliation{\textcolor{black}{Hungarian Research Network, Centre for Energy Research, Institute of Technical Physics and Materials Science, 1121 Budapest, Hungary}}

\date{\today}% It is always \today, today,

\begin{abstract}
Rhombohedral or ABC stacked multilayer graphene hosts a correlated magnetic ground state at charge neutrality, making it one of the simplest systems to investigate strong electronic correlations.
We investigate this ground state in multilayer graphene structures using the Hubbard model in a distance dependent Slater-Koster tight binding framework.
We show that by using a universal Hubbard-$U$ term, we can accurately capture the spin polarization predicted by hybrid density functional theory calculations for both hexagonal (ABA) and rhombohedral (ABC) stackings.
Using this $U$ value, we calculate the magnetic moments of 3-8 layers of ABC and ABA graphene multilayers. We demonstrate that the structure and magnitude of these magnetic moments are robust when heterostructures are built from varying numbers of ABC and ABA multilayers. By applying different types of mechanical distortions, we study the behaviour of the magnetism in graphene systems under uniaxial strain and pressure. 
Our results establish a computationally efficient framework to investigate correlation-driven magnetism across arbitrary stacking configurations of graphite polytypes.
%\begin{description}
%\end{description}
\end{abstract}

\maketitle
\
%\tableofcontents

\section{\label{sec:intro}Introduction}

Two-dimensional (2D) materials have become a central topic in solid-state physics due to their distinct properties~\cite{2D}.
Compared with three-dimensional systems, they offer advantages such as tunability by electric fields and proximity coupling to other van der Waals materials, which can reshape their band structure and interactions~\cite{num_of_papers,Kennes2021-ql}.
Few-layer graphene provides perhaps the simplest examples, including twisted bilayers~\cite{Andrei2020-hh} and rhombohedral, ABC stacked multilayers~\cite{Zhang2011-wu,abc1,abc2,abc3,abc4}, where emergent phases such as superconductivity and unconventional magnetism appear.
Lattice defects strongly influence the electronic properties of 2D materials~\cite{intro_defec1,intro_defect2,defect1,defect2,defect3,defect4,defect5}.
In multilayer systems, an additional category of defects emerges in the form of stacking faults, which arise from deviations in the regular layer-by-layer arrangement. The significance of stacking order is particularly pronounced in van der Waals layered materials, where it dramatically alters electronic properties~\cite{2d_vdw,stacking_vdw}.
Such changes in the interlayer stacking can be harnessed to tune the properties of heterostructures, a prime example being twisted structures~\cite{Andrei2020-hh,Kennes2021-ql}. 
Multilayer graphene serves as one of the simplest examples of this phenomenon.
When a graphene layer is placed on top of graphite surface with at least two layers, it can adopt two thermodynamically stable stacking configurations: the hexagonal (ABA) and the rhombohedral (ABC) positions~\cite{Nery2020-sl,Nery2021-vr}.
This pattern repeats with each additional graphene layer, leading to an exponentially growing number of possible stackings as the layer count increases~\cite{stackfault}.
Graphite in the bulk typically exhibits ABA (Bernal) stacking, while a small portion adopts ABC (rhombohedral) stacking~\cite{aba_vs_abc}.
Following this pattern, four-layer graphene should display either ABAB or ABCA sequences. However, a third type: ABCB has been experimentally observed, showing ferroelectric properties not present in the defect free structures~\cite{stackfault}.

Graphene multilayers with ABC stacking sequence, have recently gained prominence as a platform for probing emergent, strongly-correlated electronic phenomena~\cite{abc1,abc2,abc3,abc4}. Both the lattice defects and electronic properties of this material have started to be investigated~\cite{graphene_defect1,graphene_defect2,stackfault,Nery2020-sl,Nery2021-vr,Kai2025-ft}. Various theoretical approaches have been employed, ranging from simple continuum~\cite{continuum1,continuum2} and tight-binding (TB) models with parameters fitted to either experiments~\cite{tb_exp1,tb_exp2,tb_exp3} or first-principle calculations~\cite{tb_dft1,tb_dft2}. The most advanced techniques, including the GW approximation~\cite{gw,rpa}, density functional theory (DFT) using hybrid functionals~\cite{abc_hse,aba_hse}, and random phase approximation (RPA)~\cite{rpa}, have also been applied to few-layer systems. As theoretical methods become more accurate and precise, computational costs increase significantly. Calculations employing hybrid functionals demand substantial high-performance computing (HPC) resources and considerable time even on advanced systems. Conversely, TB calculations offer analytically tractable or computationally inexpensive approaches to investigate electronic structure, making them capable of modelling various lattice defects in large-scale systems containing tens of thousands of atoms~\cite{largescale1,largescale2,largescale3}.
While TB calculations offer computational efficiency, they traditionally neglect electron-electron interactions, which can be crucial for understanding many-body phenomena in graphene systems. To address this limitation, we apply the methods suggested by Hubbard
to include electron-electron interactions in the TB Hamiltonian, enabling the description of conductor-insulator transitions~\cite{hubbard_theory}. This model incorporates repulsive contact interactions parametrized by the so-called Hubbard-$U$ term, with the resulting Hamiltonian often called the TB+U Hamiltonian. This $U$ is a free parameter in the model and its value should be fitted to experimental results or \textit{ab initio} calculations.

In this work, we propose a parametrization of the TB Hamiltonian of multilayer graphene systems, using the distance-dependent Slater-Koster parametrization and extend it by adding the Hubbard term.
We show that a universal Hubbard-$U$ value can be used to describe the magnetic properties of multilayer graphene systems with arbitrary stacking configurations.
Although ABC graphite is predicted to host several competing many-body ground states~\cite{Zhou2021-wf,Braz2024-oc,Zhang2011-wu}, we focus on the zero-doping case without external electric or magnetic fields.
In this regime, experiments consistently report an insulating ground state at charge neutrality in pristine ABC-stacked samples~\cite{Lee2014-zc,Liu2024-mc,Liu2024-gh,Han2024-eb}.
This insulating phase, known as the layer antiferromagnet (LAF)~\cite{Lee2014-zc}, is characterized by antiparallel spin alignment between the outermost layers of the ABC ladder.
The LAF state matches \emph{ab initio} predictions of a magnetic insulator~\cite{abc_hse} and has been observed in tetralayer ABC~\cite{Liu2024-mc,Liu2024-gh} and pentalayer samples~\cite{Han2024-eb} among others, with possible beyond-mean-field corrections in thicker systems~\cite{Hagymasi2022-hg,Shi2020-bv}.
Here we investigate how the magnetic moments of the LAF phase are modified as the stacking changes from ABC to ABA.
More generally, our framework can also describe complex stacking geometries, including twisted multilayers~\cite{Waters2023-pe} and lateral domain walls between ABC and ABA regions, which have been detected in scanning probe experiments~\cite{Yin2017-va,Seifert2024-kd}.

\begin{figure*}
\includegraphics[scale=.5]{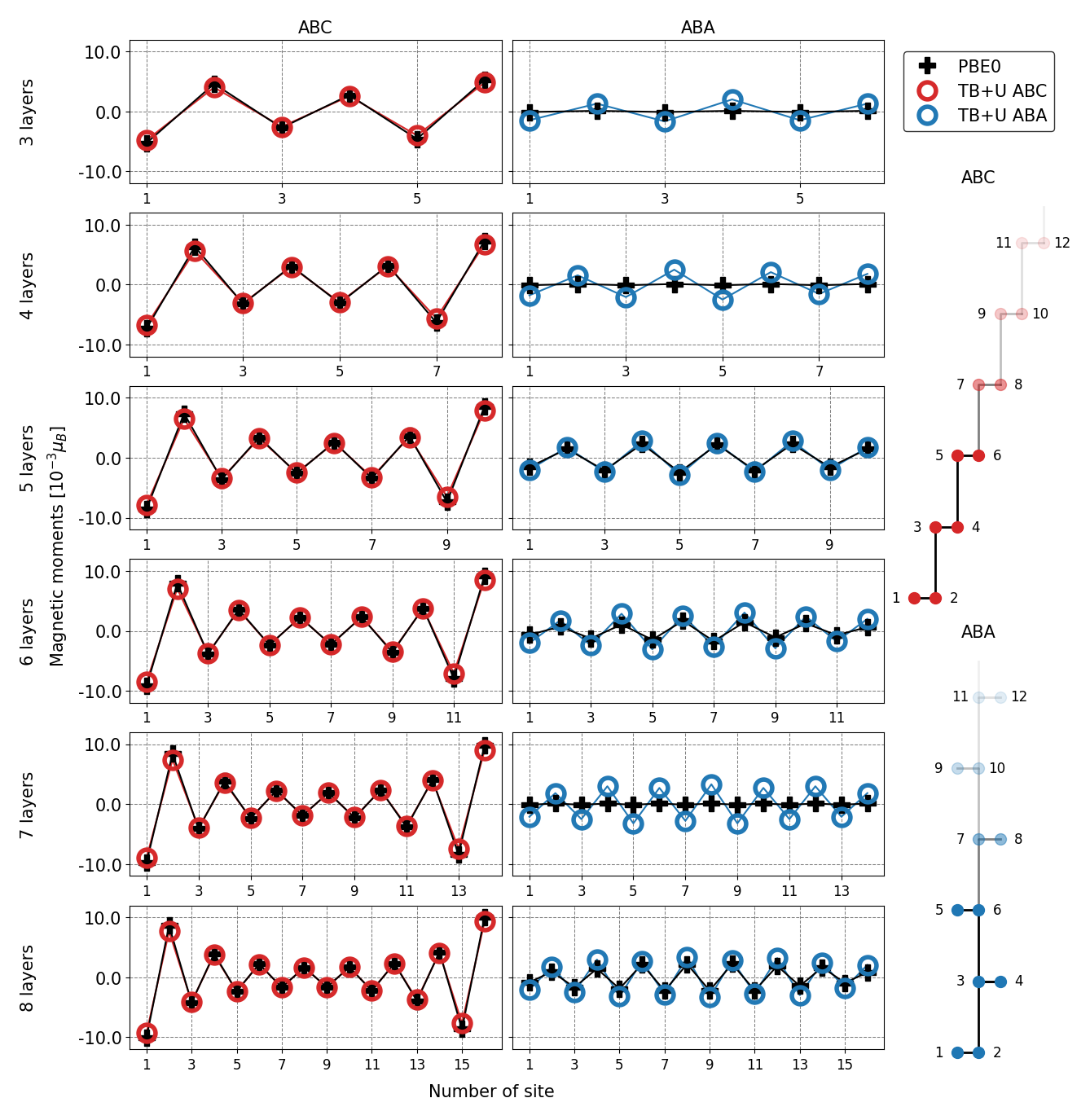}
\caption{
\label{fig:mag_moms_fit} Left column: Comparison between the magnetic moments obtained from the TB+U calculations (red) and the data from Ref.~\cite{abc_hse} (black).
Middle column:  Comparison between the magnetic moments obtained from the TB+U calculations (blue) and the data from Ref.~\cite{aba_hse} (black).
Right column: The structure and numbering of sites in ABC (top) and ABA (bottom) graphene
}
\end{figure*}

\section{Model and methods}
To investigate the magnetic properties of multilayer graphene systems, we first focus on the two most common stacking arrangements: ABA (Bernal) and ABC (rhombohedral) stacking. These structures were modeled using the following lattice vectors:
\begin{equation}
    \mathbf{a}_1=\frac{a_{\mathrm{CC}}}{2}
    \begin{pmatrix}
    3\\ \sqrt{3}
    \end{pmatrix},\ \ \ \ \ \ \ \mathbf{a}_2=\frac{a_{\mathrm{CC}}}{2}
    \begin{pmatrix}
        3 \\ -\sqrt{3}
    \end{pmatrix}
    \label{eq:lat_vecs}
\end{equation}
Where $a_{\mathrm{CC}}=a_0/\sqrt{3}$ is the carbon-carbon distance used in our work, and $a_0=2.461$~\AA\ is the in-plane lattice constant. The interlayer distance was set to $d_0=3.347$~\AA. 

These lattice parameters were chosen to be consistent with the values used in previous PBE0 calculations~\cite{abc_hse,aba_hse}.
The two stacking arrangements differ in their interlayer registry. In ABA graphene, the second layer is shifted by the vector $a_{\mathrm{CC}}(1,0)$ relative to the first layer, while the third layer is positioned directly above the first, and the fourth above the second. In ABC stacking, the first and second layers maintain the same relative positioning as in ABA, but the third layer is shifted by $2a_{\mathrm{CC}}(1,0)$ relative to the first layer. The unit cell structures of both systems are illustrated in the right column of Fig.~\ref{fig:mag_moms_fit}, where even (odd) numbered sites indicate the A (B) sublattices within each layer.

To investigate the robustness of the magnetic structure in ABC graphene, we studied two types of systems consisting of both ABC and ABA parts, as simple examples for mixed stackings.
The first type consists of an $M$-layer ABC region stacked on top of an $N$-layer ABA region. The second type features an $N$-layer ABA region sandwiched between two ABC regions of thickness $M_1$and $M_2$ layers, respectively. Since these mixed stackings are composed of alternating rhombohedral (ABC) and Bernal (ABA) multilayer segments, we adopt the naming convention R$M_1$-B$N$-R$M_2$, where R denotes rhombohedral stacking, B denotes Bernal stacking, and the subscripts indicate the number of layers in each region.
For structures with only two regions, the convention simplifies to R$M$-B$N$. Using this notation, the two example structures shown in Fig.~\ref{fig:heterostructs} are designated as R7-B10 and R5-B6-R5.
Our approach combines tight-binding (TB) calculations with Hubbard interaction terms to model the magnetic properties of multilayer graphene systems. To parametrize and validate this TB+$U$ model, we utilize reference data from recent hybrid density functional theory (DFT) calculations. Specifically, we employ the magnetic moments calculated using PBE0 hybrid functionals by Pamuk et al.~\cite{abc_hse} for ABC multilayers and Campatella et al.~\cite{aba_hse} for ABA multilayers. Both studies systematically investigated 3-8 layer systems and examined the effects of different basis sets and functionals on the electronic and magnetic structure.
The TB+$U$ model employs a tight-binding Hamiltonian extended with on-site Hubbard interaction terms to account for electron-electron repulsion. The Hubbard-$U$ parameter was determined by fitting the calculated magnetic moments to the hybrid DFT results from the aforementioned studies. The detailed formulation of the TB+$U$ model, the fitting procedure, and the computational implementation are described in \ref{sec:tb}, \ref{sec:fit} and \ref{sec:TBU}.

\section{Results}

We now examine the magnetic properties of different multilayer graphene systems using our TB+$U$ model with the fitted Hubbard parameter $U=5.84\text{ eV}$. We begin with pure stacking configurations before investigating the behaviour of mixed stacking systems. The magnetic moments of ABC multilayers calculated from our TB+$U$ model show agreement with the hybrid DFT results, as demonstrated in Fig.~\ref{fig:mag_moms_fit}. The magnetic structure exhibits a consistent pattern across all layer numbers: within each layer, the system displays antiferromagnetic ordering, with magnetic moment magnitudes that decay toward the center of the multilayer. Additionally, the overall magnetic moments increase systematically with the number of layers. \textcolor{black}{The value of the magnetic moments from our TB+U calculations and~\cite{abc_hse} are reported in Table \ref{tab:magnetic_moments_abc}. }
For systems with more than 12 layers the change of the maximum value of the magnetic moments between consecutive layer number is smaller than the convergence criteria.

The ABA multilayers exhibit fundamentally different magnetic behaviour compared to their ABC counterparts. While they also show antiferromagnetic ordering within each layer (Fig.~\ref{fig:mag_moms_fit}), the spatial distribution of magnetic moments is inverted: the magnitude increases as we move toward the center of the system. Notably, although the hybrid functional calculations predict no magnetic moments for three, four, and seven layers, our TB+$U$ calculations reveal non-zero magnetic moments for these systems as well.
Importantly, a single Hubbard parameter adequately describes the magnetic properties across both stacking configurations and multiple layer thicknesses (3-8 layers), suggesting reasonable transferability of our TB+$U$ approach. \textcolor{black}{The value of the magnetic moments from our work and~\cite{aba_hse} are reported in Table \ref{tab:magnetic_moments_aba}. }
\begin{table}
\begingroup
\arrayrulecolor{black}
\color{black}
\caption{\label{tab:magnetic_moments_abc}\textcolor{black}{The magnetic moments from our calculations and Ref.~\cite{abc_hse} marked with $^{\dagger}$ in units of $10^{-3}\mu_B$ in the ABC multilayers. Due to symmetry reasons magnetic moment $m_i=-m_{2N-i+1}$, so we only report magnetic moments up to $m_N$}}
\begin{ruledtabular}
\begin{tabular}{c d{2} d{2} d{2} d{2} d{2} d{2} d{2} d{2}}
N&  \multicolumn{1}{c}{$m_1$}
    & \multicolumn{1}{c}{$m_2$}
    & \multicolumn{1}{c}{$m_3$}
    & \multicolumn{1}{c}{$m_4$}
    & \multicolumn{1}{c}{$m_5$}
    & \multicolumn{1}{c}{$m_6$}
    & \multicolumn{1}{c}{$m_7$}
    & \multicolumn{1}{c}{$m_8$} \\ \hline
\multirow{2}{*}{3} 
& -4.80 & 4.01 & -2.55&  &  &  &  & \\ 
& -5.28^{\dagger}& 4.60^{\dagger} & -2.73^{\dagger} &  &  &  &  & \\ 
\multirow{2}{*}{4} 
& -6.76 & 5.61 & -3.10& 2.96 &  &  &  & \\ 
& -7.32^{\dagger}&  6.33^{\dagger}& -3.22^{\dagger} &3.11^{\dagger}  &  &  &  & \\ 
\multirow{2}{*}{5} 
& -7.86 & 6.51& -3.46 & 3.23 & -2.50 &  &  & \\ 
& -8.56^{\dagger}& 7.38^{\dagger} & -3.62^{\dagger} & 3.43^{\dagger} &-2.60^{\dagger}    &  &  & \\ 
\multirow{2}{*}{6} 
& -8.55 & 7.06& -3.73& 3.44 & -2.34 & 2.28 &  & \\ 
& -9.19^{\dagger}& 7.90^{\dagger} & -3.82^{\dagger} & 3.58^{\dagger} & -2.34^{\dagger} & 2.29^{\dagger} &  & \\ 
\multirow{2}{*}{7} 
& -9.00 & 7.42 & -3.94 & 3.61 & -2.31 & 2.21 & -1.87 & \\ 
& -9.75^{\dagger}& 8.40^{\dagger} & -4.11^{\dagger} & 3.84^{\dagger} & -2.38^{\dagger} & 2.30^{\dagger} & -1.93^{\dagger} & \\ 
\multirow{2}{*}{8} 
& -9.31 & 7.68 & -4.10 & 3.74 & -2.32 & 2.20 & -1.69 & 1.66\\ 
& -10.04^{\dagger} & 8.64^{\dagger} & -4.23^{\dagger} & 3.94^{\dagger} & -2.34^{\dagger} & 2.24^{\dagger} & -1.69^{\dagger} & 1.66^{\dagger}\\ 
\end{tabular}
\end{ruledtabular}
\endgroup
\end{table}

\begin{table}
\color{black}
\caption{\label{tab:magnetic_moments_aba}\textcolor{black}{The magnetic moments from our calculations and Ref.~\cite{aba_hse}, marked with $^{\dagger}$ in units of $10^{-3}\mu_B$ in the ABA multilayers. Due to symmetry reasons magnetic moments are reported only up to $m_N$ for systems with even number of layers since $m_i=-m_{2N-i+1}$. For systems with odd number of layers we report up to $m_{N+1}$}. For layer numbers 3,4 and 7 Ref.~\cite{aba_hse} reports negligible magnetic moments ($<0.1\cdot 10^{-3}\mu_B$), therefore we indicate the reference values as $<$0.1 in the table.}
\begin{ruledtabular}
\begin{tabular}{c  d{2} d{2} d{2} d{2} d{2} d{2} d{2} d{2}}
N & \multicolumn{1}{c}{$m_1$}
    & \multicolumn{1}{c}{$m_2$}
    & \multicolumn{1}{c}{$m_3$}
    & \multicolumn{1}{c}{$m_4$}
    & \multicolumn{1}{c}{$m_5$}
    & \multicolumn{1}{c}{$m_6$}
    & \multicolumn{1}{c}{$m_7$}
    & \multicolumn{1}{c}{$m_8$} \\ \hline
\multirow{2}{*}{3} 
& 1.47 & -1.30 & 1.66 & -2.00 &  &  & \\ 
& <0.1^{\dagger} & <0.1^{\dagger} & <0.1^{\dagger} & <0.1^{\dagger} &  &  &  & \\ 
\multirow{2}{*}{4} 
& 1.79 & -1.55 & 2.08 & -2.49 &  &  & \\ 
& <0.1^{\dagger} & <0.1^{\dagger} & <0.1^{\dagger} & <0.1^{\dagger} &  &  &  & \\ 
\multirow{2}{*}{5} 
& 1.99 & -1.74 & 2.34 & -2.80 & 2.88 & -2.44 & & \\ 
& 1.56^{\dagger} & -1.48^{\dagger} & 2.10^{\dagger} & -2.29^{\dagger} & 2.40^{\dagger} & -2.22^{\dagger} & &\\ 
\multirow{2}{*}{6} 
& 2.01 & -1.75 & 2.40 & -2.87 & 3.06 & -2.57 & &\\ 
& 0.65^{\dagger} & -0.71^{\dagger} & 1.26^{\dagger} & -1.08^{\dagger} & 1.47^{\dagger} & -1.70^{\dagger} & & \\ 
\multirow{2}{*}{7} 
& 2.10 & -1.84 & 2.50 & -2.99 & 3.20 & -2.71 & 2.78 & -3.31 \\ 
& <0.1^{\dagger} & <0.1^{\dagger} & <0.1^{\dagger} & <0.1^{\dagger} & <0.1^{\dagger} & <0.1^{\dagger} & <0.1^{\dagger} & <0.1^{\dagger}\\ 
\multirow{2}{*}{8} 
& 2.07 & -1.80 & 2.48 & -2.97 & 3.21 & -2.70 & 2.82 & -3.35\\ 
& 0.83^{\dagger} & -0.93^{\dagger} & 1.67^{\dagger} & -1.44^{\dagger} & 1.90^{\dagger} & -2.20^{\dagger} & 2.24^{\dagger} & -2.21^{\dagger}\\ 
\end{tabular}
\end{ruledtabular}
\end{table}

\begin{figure*}
   \centering
    \includegraphics[scale=.5]{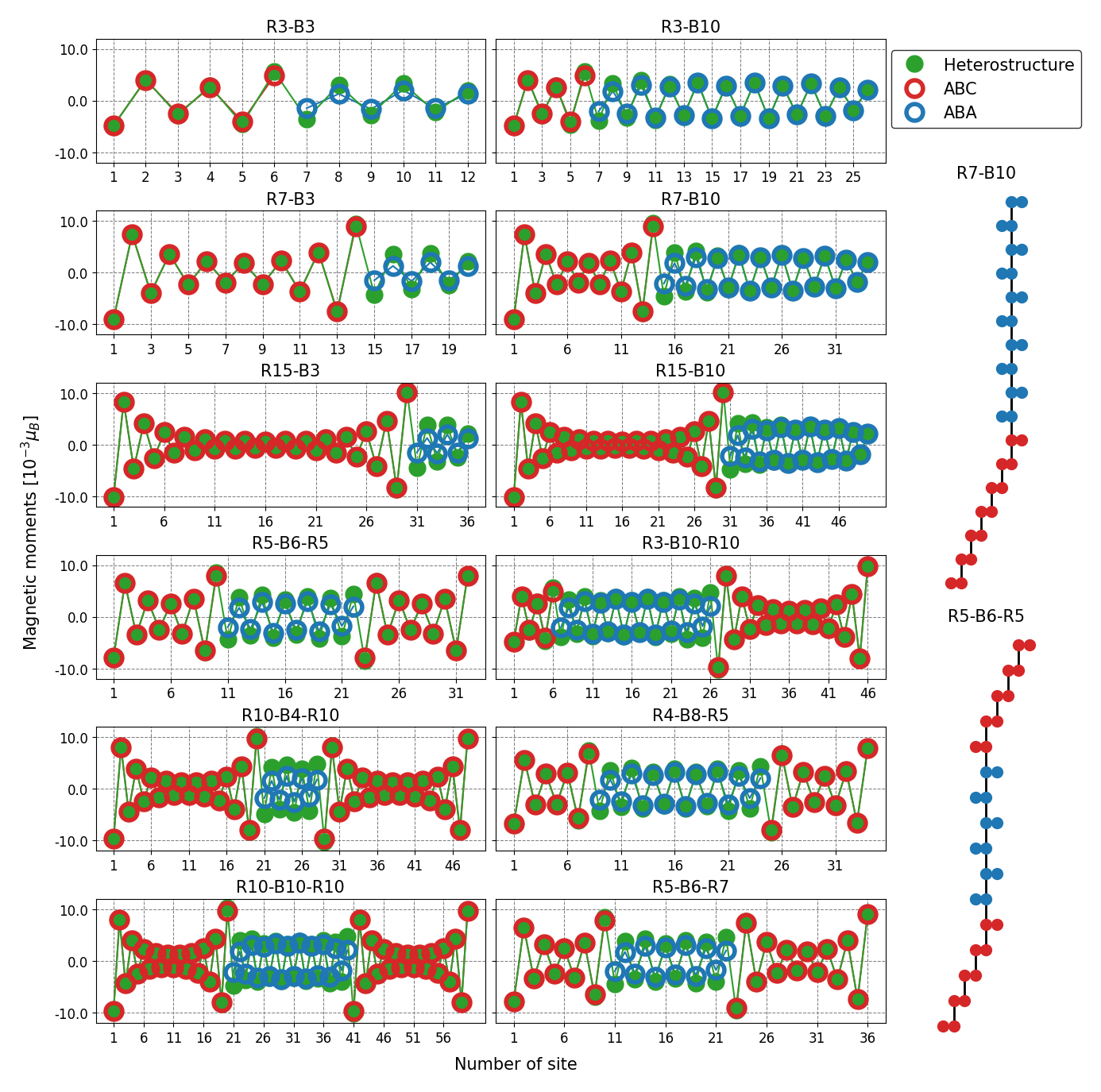}
    \caption{
    The magnetic structure and geometry of various ABC-ABA mixed stackingss. Empty red (blue) circles represent the magnetic moments for ABC (ABA) graphene multilayers of the same length. Filled green circles represent the magnetic structure of the whole mixed stackings.
    }
    \label{fig:heterostructs}
\end{figure*}

Having validated our model against pure systems, we now turn to ABC-ABA mixed stackings, which represent a computationally prohibitive challenge for hybrid DFT methods due to their large system and configurational sizes. 
To investigate how the distinct magnetic behaviours of ABC and ABA regions interact, we calculated the magnetic structure of various ABC-ABA mixed stackings shown in Fig.~\ref{fig:heterostructs}. The results reveal that the magnetic moment patterns are largely preserved within each stacking region, with the ABC portions maintaining their characteristic decay toward the center while the ABA regions exhibit their typical center-enhanced behaviour.
In the ABA regions located far from the ABC-ABA junction (typically 2-3 layers away), the magnetic structure remains virtually indistinguishable from that of a free-standing ABA system. However, as the number of ABC layers in the mixed stackings increases, the perturbation to the ABA magnetic structure becomes more pronounced, indicating a non-local influence of the ABC stacking on the overall magnetic properties.

\subsection{Effects of mechanical distortion}
Since the extent of the surface flat band, that gives rise to the magnetic instability is proportional to the interlayer hopping terms~\cite{continuum1}, it is expected that the magnetic moments in the LAF phase should scale with the interlayer hopping term.
Having established the magnetic properties of unstrained systems, we now investigate how mechanical deformation affects the magnetic structure of multilayer graphene.

\subsubsection{Uniaxial strain in the armchair direction}
\begin{figure}[b]
    \centering
    \includegraphics[scale=.4]{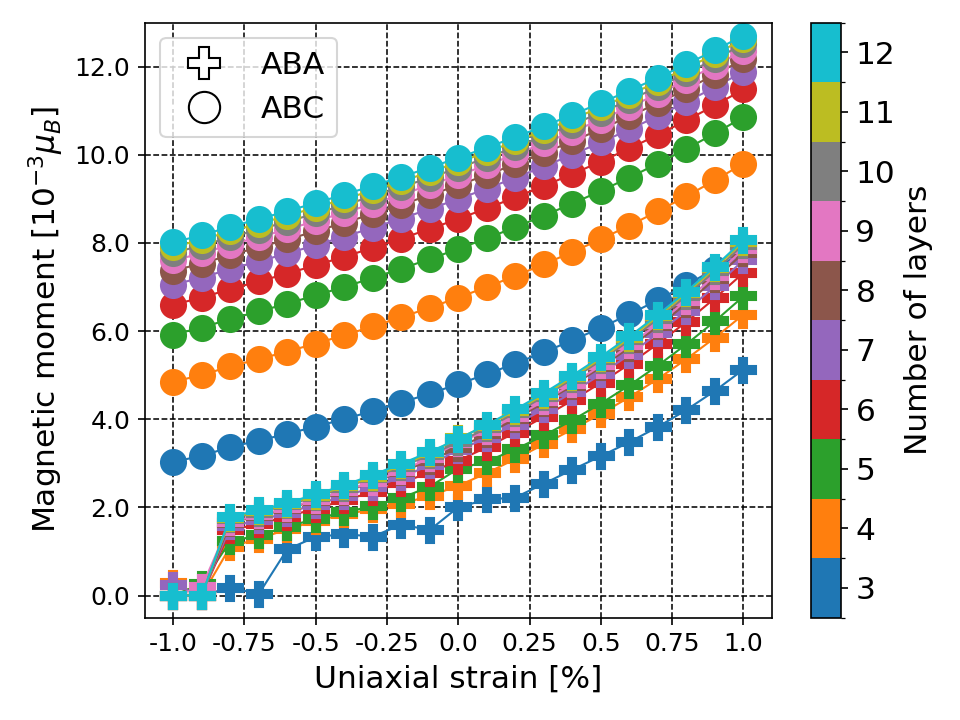}
    \caption{The value of the maximum magnetic moment in 3-12 layer ABC and ABA multilayers in the case of uniaxial strain along the armchair direction.}
    \label{fig:uniaxial}
\end{figure}

We applied uniaxial strain to study its effects on the magnetic structure of pure ABC and ABA multilayers as well as selected mixed stackings. The strain was applied in the armchair ($x$) direction, parallel to one of the carbon-carbon bonds, with magnitudes of $\pm 1\%$. \textcolor{black}{Due to graphene's physical properties' insensitivity to the direction of strain~\cite{armchair_vs_zigzag_strain} in the case of small distortions, we only study uniaxial strain in the armchair direction.}
%For the mixed stacking analysis, we selected two representative systems that cover different structural characteristics: R6-B3 and R7-B7-R7. The R6-B3 mixed stacking contains a single ABC-ABA junction with a relatively thin ABA region, while the R7-B7-R7 structure features two junctions with symmetric ABC regions flanking a central ABA layer. This selection allows us to probe the strain response across a range of junction geometries and layer distributions.
The calculations were performed on multilayer systems with \textcolor{black}{N=3-12} for both pure ABC and ABA configurations. The results, shown in Fig.~\ref{fig:uniaxial}, display the maximum magnetic moment values for each system to provide a clear comparison of the strain effects across different configurations.
Under tensile strain (stretching), the magnitude of magnetic moments increases in ABC, ABA. Conversely, compressive strain exhibits asymmetric effects on the two stacking types: in ABA systems, magnetism can be suppressed by sufficiently large compression, while ABC systems maintain their magnetic character throughout the experimentally accessible strain range~\cite{exp_strain}. \textcolor{black}{Magnetism in multilayer graphene is controlled by electronic correlation effects~\cite{correlation}, which are determined by the ratio of in-plane to interlayer hopping amplitudes. The application of mechanical strain reduces the in-plane hopping parameters~\cite{armchair_vs_zigzag_strain}, thereby enhancing electronic correlations and strengthening the magnetic instability.}

\begin{figure}[b]
    \centering
    \includegraphics[scale=.4]{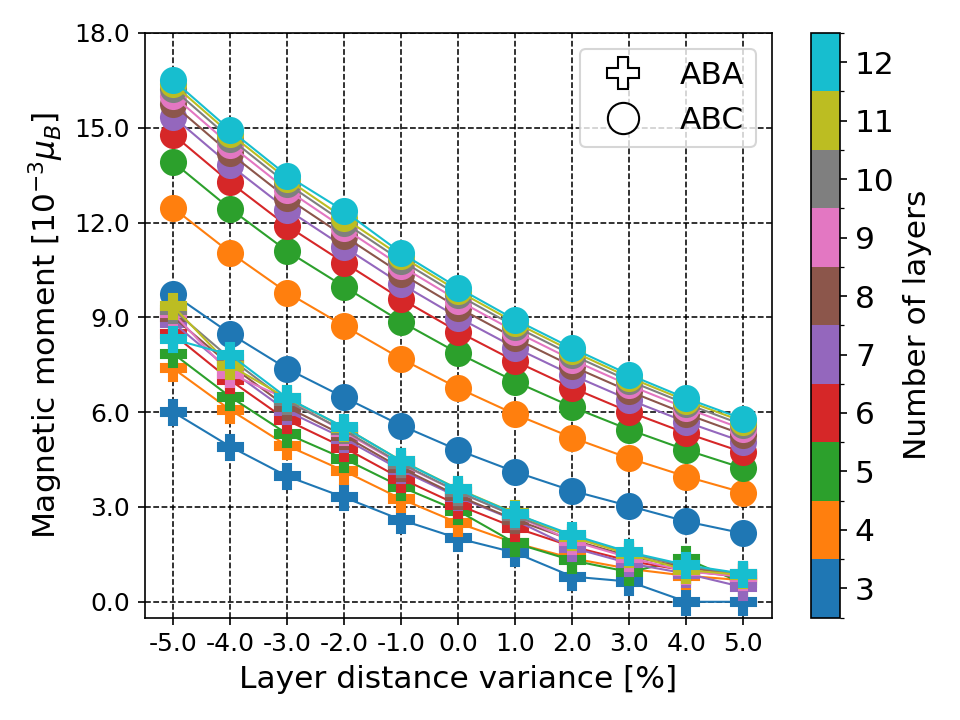}
    \caption{The value of the maximum magnetic moment in 3-12 layer ABC and ABA multilayers in the case of modified van der Waals separation.
    }
    \label{fig:z}
\end{figure}

\subsubsection{Deformation of the van der Waals spacing}

In addition to in-plane strain effects, we investigated how interlayer separation influences the magnetic properties by applying compression along the $z$ direction (perpendicular to the graphene planes). This allows us to probe the role of interlayer coupling in stabilizing the magnetic states.
We applied $\pm 5\%$ changes to the interlayer distance, where $+ 5\%$ corresponds to an increased layer separation of $1.05~d_0$ relative to the equilibrium distance $d_0 = 3.347$~\AA.
We chose higher strain values because compression values of up to 5\%\ are available experimentally in van der Waals systems when hydrostatic pressure is applied~\cite{Szentpeteri2021-px}.
The same multilayer systems \textcolor{black}{N=(3-12)} were examined for both ABC and ABA stackings. The results presented in Fig.~\ref{fig:z} show the maximum magnetic moment for each system configuration, allowing for straightforward comparison of how interlayer distance affects magnetic strength across different layer numbers and stacking types.

The calculations reveal a universal trend across all layer numbers and stacking configurations: increasing the interlayer distance consistently reduces the magnitude of magnetic moments, while decreasing the separation enhances them. This behaviour can be understood through the weakening of interlayer orbital interactions as the layers are moved further apart. The reduced overlap between $\pi$ orbitals in adjacent layers diminishes the electronic coupling that drives the magnetic instability, leading to smaller magnetic moments.
Conversely, when layers are brought closer together, the enhanced interlayer interactions strengthen the conditions for magnetic ordering. This $z$-direction response demonstrates that the magnetic properties of multilayer graphene are sensitive not only to in-plane deformation but also to the three-dimensional structural parameters.

\section{Conclusion and outlook}
We have shown that a universal Hubbard-$U$ value can be used to capture the magnetic properties predicted by hybrid DFT calculations in defect free ABA and ABC stacked graphene multilayers. Our calculations showed that the magnetic structure of ABC graphene is robust when it is combined with ABA graphene into structures with mixed stacking. 
At the ABC–ABA interface, the magnetic moments within two layers of the boundary are reduced compared to those in free-standing ABA graphene, but farther away they converge to the same values.
Interestingly the edge momenta of ABC regions are less perturbed and remain equal to their free-standing values when adjacent to ABA stacking.
This suggests that "buried" flat-band~\cite{Garcia-Ruiz2023-tu} magnetic states could be prepared, where neighbouring hexagonal regions can protect the magnetism from charge disorder.

To incorporate the effect of different mechanical distortions we applied uniaxial strain in the armchair direction and pressure perpendicular to the graphene planes. By decreasing the layer distance, the orbitals within adjacent layers get closer to each other, therefore the interaction gets larger between the electrons. This causes the magnetic moments on each site to increase. On the other hand calculations with uniaxial strain showed opposite behaviour in the magnetic structure. By stretching the system the magnetic moments become larger, while by compressing it they become smaller. Based on the results from the calculations incorporating both strain and pressure, the absence of magnetism in 3, 4 and 7 layer ABA graphene in Ref.~\cite{aba_hse}, could be explained with geometrical reasons, for details see Appendix \ref{sec:xz_heatmap}. 

The emergence of magnetic moments in the hexagonal phase, which our calculations also support is a noteworthy and often overlooked feature.
This should motivate future experimental work and many-body calculations, since charge transport measurements already suggest signatures of correlation effects~\cite{Nam2018-da}.

\begin{acknowledgments}
Supported by the DKOP-23 Doctoral Excellence Program of the Ministry for Culture and Innovation from the source of the National Research, Development and Innovation Fund. This work was supported by the National Research, Development and Innovation Office (NKFIH) in Hungary, through Grant No. FK-142985, K 146156 and Excellence 151372 and by the Hungarian Academy of Sciences LP2024-17 Lend\"{u}let "Momentum" grant. Z.T. acknowledges support from the János Bolyai Research Scholarship of the Hungarian Academy of Sciences. This project is supported by the TRILMAX Horizon Europe consortium (Grant No. 101159646). We acknowledge the Digital Government Development and Project Management Ltd. for awarding us access to the Komondor HPC facility based in Hungary.
\end{acknowledgments}
\newpage
\appendix

\section{The tight-binding model}

\subsection{The model}
\label{sec:tb}
To determine the electronic properties of the systems a distance dependent tight-binding model was used. The tight-binding Hamiltonian for electrons in a single sheet of graphene considering that electrons can hop between first, second and third nearest neighbours has the following form:
\begin{widetext}
\begin{equation}
\hat{\mathrm{H}}_{\mathrm{TB},\mathrm{in-plane}}=-\gamma_1\sum_{\langle i,j \rangle,\sigma}\left(a^{\dagger}_{i,\sigma}b_{j,\sigma} + \mathrm{H.c.}\right) -\gamma_2\sum_{\langle\langle i,j \rangle\rangle,\sigma}\left(a^{\dagger}_{i,\sigma}a_{j,\sigma} + b^{\dagger}_{i,\sigma}b_{j,\sigma}+ \mathrm{H.c.}\right)-\gamma_3\sum_{\langle\langle\langle i,j \rangle\rangle\rangle,\sigma}\left(a^{\dagger}_{i,\sigma}b_{j,\sigma} + \mathrm{H.c.}\right)
\label{eq:TB_Ham_inplane}
\end{equation}
\end{widetext}
The operators $a^{\dagger}_{i,\sigma}\ (a_{i,\sigma})$ create (annihilate) an electron on sublattice A, site $i$ with spin $\sigma$. The notation is the same for the $b$ operators. $\langle\cdots\rangle$ denotes summing over the first-, $\langle\langle\cdots\rangle\rangle$ denotes summing over the second- and $\langle\langle\langle\cdots\rangle\rangle\rangle$ denotes summing over the third-nearest neighbours. The hoppings have the following form~\cite{gorog}:
\begin{equation}
\gamma_i(r) = \gamma_{i,0}\cdot \mathrm{e}^{\beta_i\left(\frac{r}{a_{\mathrm{CC}}}-1\right)}
\label{eq:inplane_hopping}
\end{equation}
where the subscript $i=1,2,3$ denotes the first, second and third neighbours, $\gamma_{i,0}$ is the value of the hopping integral in the equilibrium position, $\beta_i$ is the strength of the decay of the hopping integral and $r$ is the distance between the two atoms. Both $\gamma_{i,0}$ and $\beta_i$ were obtained from fitting the tight-binding band structure to DFT results using the least-squares method (see Sec.~\ref{sec:fit}).
To include the hoppings between the different layers, the Hamiltonian \ref{eq:TB_Ham_inplane} was extended with the following terms:
\begin{equation}
\hat{\mathrm{H}}_{\mathrm{TB},\mathrm{interlayer}}=\sum_{i,j,\sigma}\gamma_{ij}c^{\dagger}_{i,\sigma}c_{j,\sigma}+\mathrm{H.c.}
\end{equation}
The operators $c^{\dagger}_{i,\sigma},\ (c_{i,\sigma})$ create (annihilate) an electron on site $i$ with spin $\sigma$. Sites $i$ and $j$ are in adjacent layers. The interlayer hopping between atoms at positions $\mathbf{r}_i$ and $\mathbf{r}_j$ has the following form following the work of Slater and Koster~\cite{Slater_Koster}:
\begin{widetext}
\begin{equation}
\gamma_{ij}=\sum_i\left[\left(\Tilde{V}_{pp\pi}\cdot \cos^2(\theta_i)+\Tilde{V}_{pp\sigma}\cdot \sin^2(\theta_i)\right)\cdot\mathrm{e}^{-\alpha\cdot r_{ij}}\cdot\Theta(\mathbf{r}_{ij})\right]
\end{equation}
\end{widetext}

\begin{figure}
    \centering
    \includegraphics[width=0.75\linewidth]{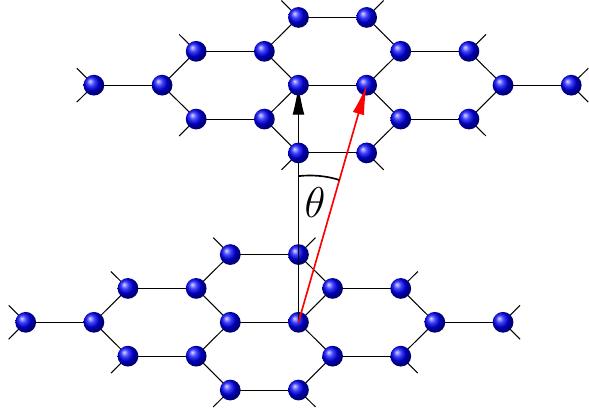}
    \caption{The $\theta_i$ angle showed in the AB bilayer. The red arrow is pointing from atom $j$ to atom $i$, while the black arrow is the $z$ axis.}
  \label{fig:theta}
\end{figure}
\begin{figure}
    \centering
    \includegraphics[width=0.75\linewidth]{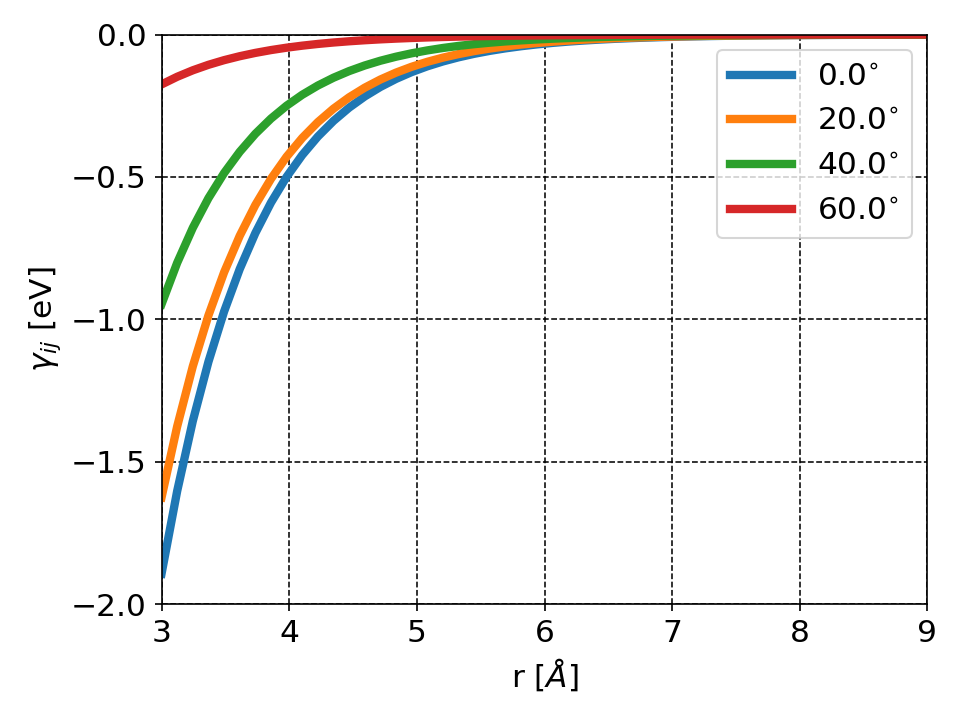}
    \caption{The distance dependence of the interlayer hopping $\gamma_{ij}$}
    \label{fig:gij}
\end{figure}

Here the angle between the vector $\mathbf{r}_{ij}=\mathbf{r}_i-\mathbf{r}_j$ and the  $z$ axis is $\theta_i$ as it can be seen in Fig.~\ref{fig:theta}. The magnitude of vector $\mathbf{r}_{ij}$ is $r_{ij}$. To eliminate the very small hoppings a cut-off function $\Theta(\mathbf{r}_{ij})$ was introduced that is 1 for $r_{ij}<7$~\AA~  and 0 otherwise. By choosing this cut-off distance we eliminate the very small matrix elements. For a few given $\theta$ values the distance dependence of the $\gamma_{ij}$ hoppings can be seen in Fig.~\ref{fig:gij}.
$\Tilde{V}_{pp\pi}$, $\Tilde{V}_{pp\sigma}$ and $\alpha$ are parameters that had to be fitted to DFT calculations.\\
\subsection{Fitting procedure}
\label{sec:fit}
\begin{figure}[b]
    \centering
    \includegraphics[width=.95\linewidth]{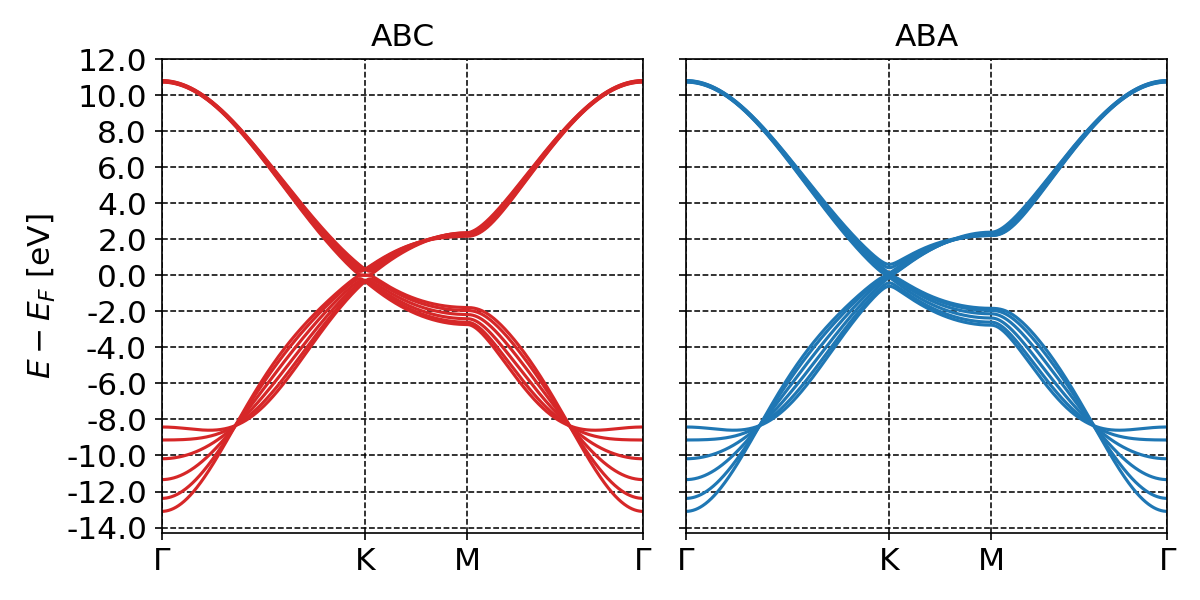}
    \caption{The band structure of 6 layer ABC and ABA graphene along the full $\Gamma-K-M-\Gamma$ high symmetry path in the Brillouin zone}
    \label{fig:full_bands}
\end{figure}

Fig.~\ref{fig:full_bands} shows the band structure of 6 layer ABC and ABA graphene along the full $\Gamma-K-M-\Gamma$ high symmetry path in the Brillouin zone. The only part of the high symmetry that has energy bands in the relevant energy range ($\pm$ 1 eV) is the small vicinity of the $K$ point. Therefore the TB parameters were obtained by fitting the band structure of the TB Hamiltonian to the dispersion relation calculated using the SIESTA code~\cite{siesta1,siesta2,siesta3} using the least squares method in a small vicinity of the $K$ point. 

\begin{figure*}
    \centering
    \includegraphics[width=0.75\linewidth]{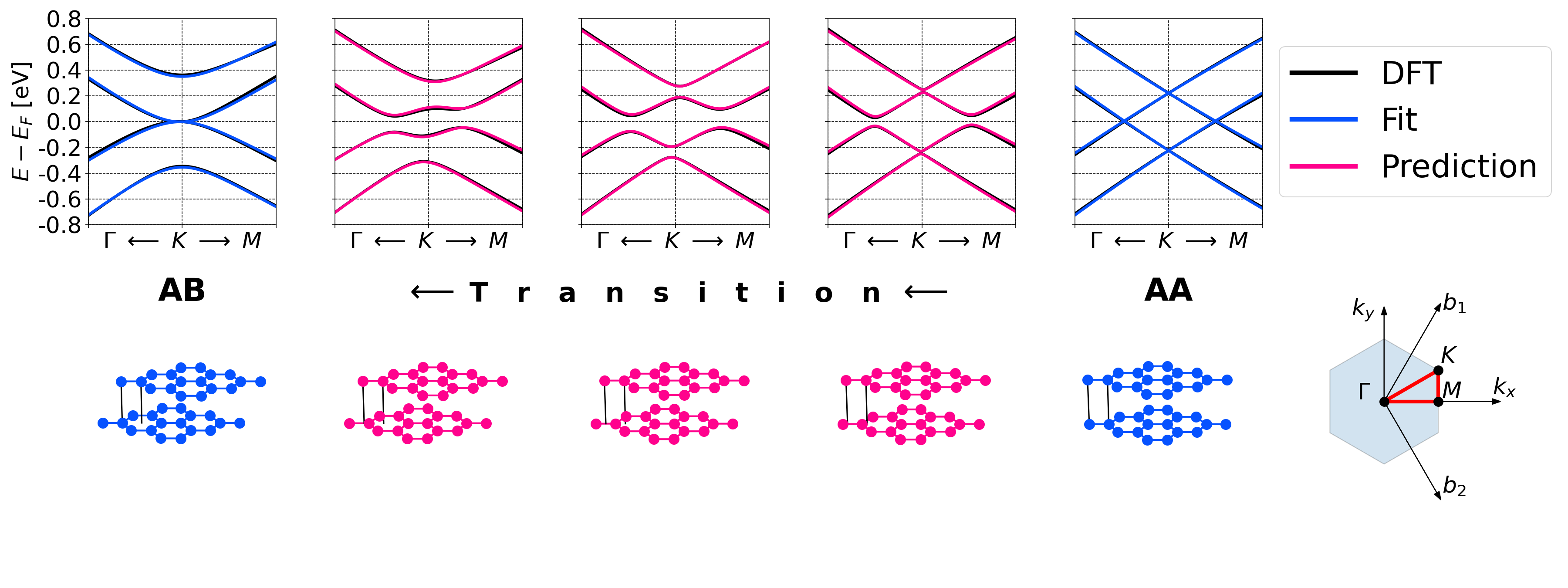}
    \caption{Top row: The DFT and TB band structures from the fitting. Bottom row: Schematic image of the geometries used in the fitting and verifying procedure. Bottom right: The high symmetry path in the Brillouin zone}
    \label{fig:fit}
\end{figure*}

First the nearest neighbour in-plane hopping were fitted to the monolayer graphene band structure. 
To include the distance dependence, biaxial strain was applied to the system, and the value of $\gamma_{1,0}$ was obtained by fitting the TB band structure to the DFT dispersion relation. After that the a function of the form of Eq.~\ref{eq:inplane_hopping} was fitted to the different hopping values to get the value of $\beta_1$. To further improve the precision of the model, the second the third nearest neighbour hoppings were introduced. The $\gamma_{2,0}$, $\gamma_{3,0}$ hopping values and their respective $\beta$ values were fitted to different DFT band structures when uniaxial strain was applied. 
Regarding the parameters describing the interlayer hoppings the DFT band structure of AB and AA bilayer graphene was used. The three parameters ($\Tilde{V}_{pp\pi}$, $\Tilde{V}_{pp\sigma}$, $\alpha$) were simultaneously fitted to the AB and AA band structure by using four low energy bands from the AB and four low energy bands from the AA band structure. Fig.~\ref{fig:fit}. shows the fitting procedure the interlayer hopping parameters.  
\subsection{TB+U}
\label{sec:TBU}
To include the electron-electron interaction we extended the Hamiltonian with the Hubbard term~\cite{hubbard_theory}. Therefore the Hamiltonian has the following form:
\begin{equation}
\hat{\mathrm{H}} = \hat{\mathrm{H}}_{\mathrm{TB}}+\sum_{i,\uparrow,\downarrow}U_in_{i\uparrow}n_{i\downarrow}
\end{equation} 
Where $\hat{\mathrm{H}}_{\mathrm{TB}}=\hat{\mathrm{H}}_{\mathrm{TB},\mathrm{interlayer}}+\hat{\mathrm{H}}_{\mathrm{TB},\mathrm{in-plane}}$. By applying the meanfield approximation, the Hamiltonian gets the following form:
\begin{equation}
    \hat{\mathrm{H}}_{\mathrm{MF}} = \hat{\mathrm{H}}_{\mathrm{TB}}+U\sum_{i,\sigma}\langle n_{i\sigma}\rangle n_{i\bar{\sigma}}
\end{equation}
Where instead of having different $U_i$ values on each site there is one global $U$ for all of them and $\langle n_{\sigma}\rangle$ is the $\sigma=\uparrow,\downarrow$ spin density on site $i$. In the above Hamiltonian the only free parameter is the Hubbard $U$. Its value was found by simultaneously fitting the magnetic moments of the five layer ABC and ABA systems obtained from the TB+U model to the results of the hybrid DFT calculations~\cite{abc_hse,aba_hse}. The above Hamiltonian contains the expectation value of the spin-resolved density operator $n_{\sigma}$. This expectation value is computed from the eigenvectors of $\hat{\mathrm{H}}_{\mathrm{MF}}$, therefore a self-consistent field (SCF) method should be applied. During the SCF calculations a $300\times 300 \times 1$ Monkhorst-Pack ~\cite{Monkhorst_Pack} Brillouin zone sampling was used. The temperature was et to be 10$^{-5}\ k_BT$.  
The TB and TB+U calculations were performed using the \textsc{sisl}\cite{sisl} and the \textsc{hubbard}\cite{hubbard_code,hubbard1,hubbard2} codes.
\subsection{\textcolor{black}{Goodness of fitting the Hubbard-$U$}}
\begin{figure}[b]
    \centering
    \includegraphics[width=.95\linewidth]{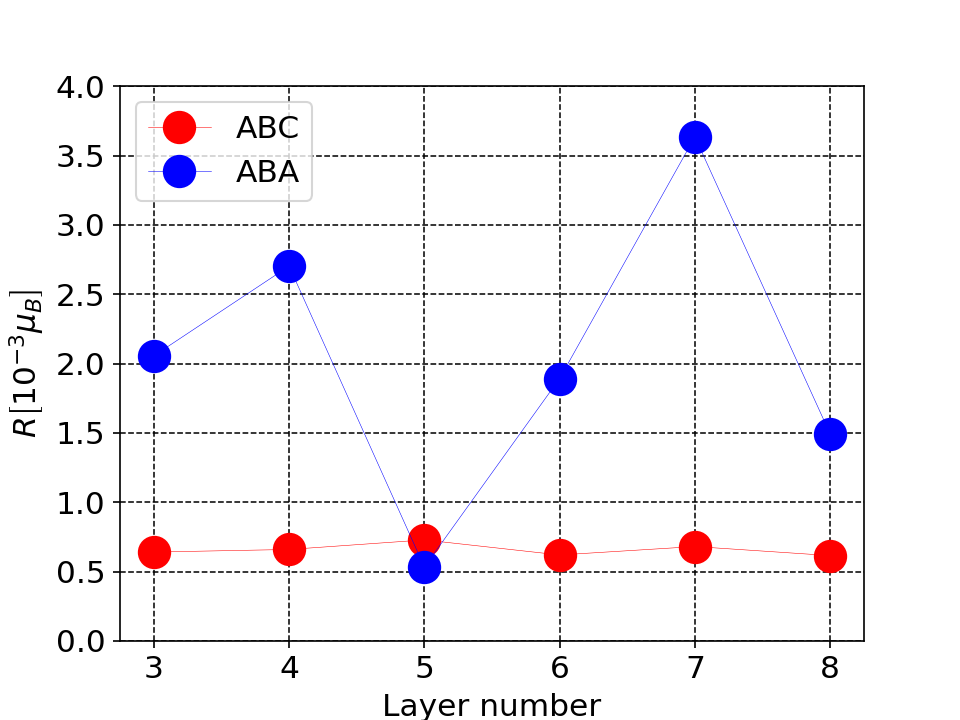}
    \caption{\textcolor{black}{The value of $R$ defined in Eq. \ref{eq:U_error} for both ABC and ABA stacking for N=3-8 layers}}
    \label{fig:U_error}
\end{figure}

\textcolor{black}{To quantify our model's power to reproduce the reference results we calculated the following expression for layer number and stacking configurations:
\begin{equation}
    R=\sqrt{\frac{1}{N}\sum_j^N(m_{\mathrm{ref},j}-m_{\mathrm{TB+U},j})^2}
    \label{eq:U_error}
\end{equation}
where $m_{\mathrm{ref}}$ are the magnetic moments from~\cite{abc_hse} and~\cite{aba_hse} and $m_{\mathrm{TB+U}}$ are the magnetic moments from our work. This $R$ value can be seen in Fig. \ref{fig:U_error}.} 
\section{Details of the DFT calculations}
The DFT calculations were performed using the SIESTA code. for the exchange-correlation functional the Perdew-Burke-Ernzerhof (PBE)~\cite{pbe} parametrization of the generalized gradient approximation (GGA) was used with double-$\zeta$ basis set. The norm-conserving pseudopotential were collected from the PseudoDojo~\cite{pseudodojo} project.\\ To ensure well converged results, a $k$-grid of $300\times 300 \times 1$ was used using the Monkhorst-Pack sampling. A real space grid cut-off energy of 500 Ry was used. The structures were relaxed until the maximum force was less than 0.008~eV/\AA. During the calculations vacuum separating the samples was chosen to be at least 30~\AA.
\section{Simultaneous uniaxial strain and compression}
\label{sec:xz_heatmap}
\begin{figure*}
    \centering
    \includegraphics[scale=.5]{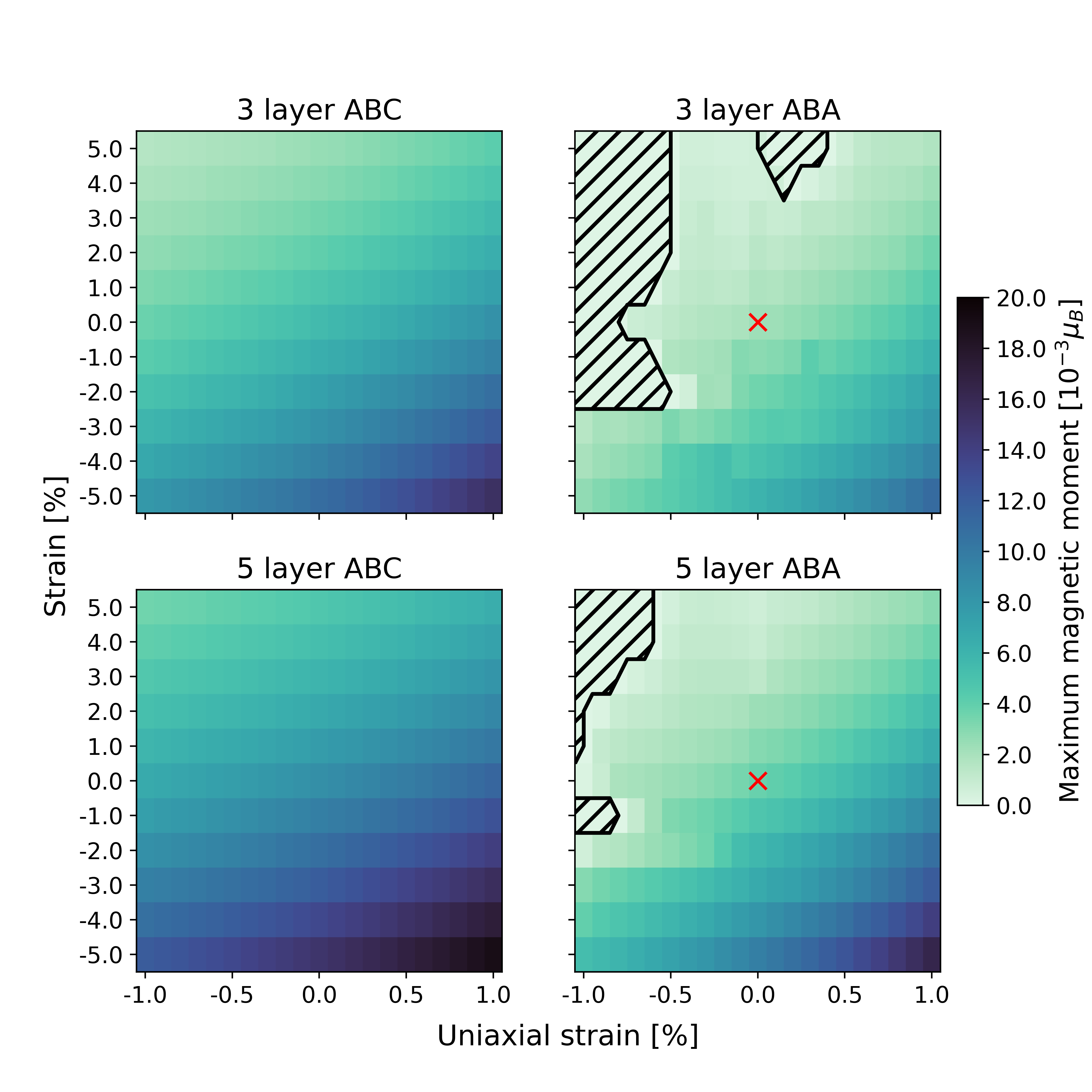}
    \caption{The value of the maximum magnetic moment is 3 and 5 layer ABC and ABA multilayers as a function of the uniaxial strain and compression. \textcolor{black}{The region where the maximum magnetic moments are smaller than 0.1$\cdot10^{-3}\mu_B$ is marked with stripes. The geometry used in Ref.~\cite{abc_hse} and \cite{aba_hse} is marked with a red cross.}}
    \label{fig:xz_heatmap}
\end{figure*}

In~\cite{abc_hse} and~\cite{aba_hse} the effect of different geometries was not studied. In accordance with this we fixed the lattice parameters. With these settings the TB+U model could not reproduce the absence of magnetism in 3, 4 and 7 layer ABA graphene. Calculation incorporating both uniaxial strain and interlayer compression (Fig.~\ref{fig:xz_heatmap}) show that for a large set of possible lattice parameters the magnetic moments are negligibly small or zero for the ABA 3 layer graphene. 
The same effect occurs in 5 layer ABA but for a much smaller set of lattice parameters. Meanwhile the magnetism is robust for both ABC multilayers. Therefore an explanation for the absence of magnetism in the 3,4 and 7 layer ABA systems in~\cite{aba_hse} is the selected values of the lattice parameters. 
\textcolor{black}{Finally, we emphasize a limitation of the present approach: our TB+$U$ model employs a single on-site (purely local) effective Hubbard parameter $U$ within a mean-field treatment. This neglects nonlocal Coulomb interaction terms (e.g., intersite interactions) and the screening/environment dependence of the effective interaction. Such beyond-local effects are expected to be most relevant when the system is close to the critical point and the magnetic moments are extremely small, as in ABA trilayer/tetralayer/heptalayer.}

%\textcolor{black}{Furthermore the Hubbard model is purely local, meanwhile DFT calculations using hybrid functionals incorporate nonlocal effects too therefore it is inevitable that the TB+$U$ model does not capture every detail of the PBE0 calculations.}

\newpage
\bibliography{apssamp}% Produces the bibliography via BibTeX.

\end{document}